\documentclass[aps,prc,superscriptaddress,showpacs,nofootinbib,floatfix,twocolumn,10pt,showkeys]{revtex4}

\usepackage{graphicx}   
\usepackage[sort&compress]{natbib}
\usepackage{ifthen}
\usepackage{hyperref} 
\usepackage{breakurl}
\usepackage{color}
\usepackage{epsfig}
\usepackage{amssymb}
\usepackage{amsmath}
\usepackage{multirow}
\newcommand{\pt}{$p_T$ }
\newcommand{\highpt}{high-\pt}
\newcommand{\be}{\begin{equation}}
\newcommand{\ee}{\end{equation}}
\newcommand{\fig}[1]{Fig.~\ref{#1}}
\newcommand{\eq}[1]{Eq.~(\ref{#1})}

\definecolor{darkgreen}{rgb}{0,.7,0}
\definecolor{linkblue}{rgb}{0.,0.,0.9333}

\newcommand{\TITLE}{A study of the correlations between jet quenching observables at RHIC}

\hypersetup{
    pdffitwindow=true,      
    pdftitle={\TITLE},    
    pdfauthor={Jiangyong Jia, W. A. Horowitz, and Jinfeng Liao},     
    pdfnewwindow=true,      
    bookmarksnumbered=true,
    colorlinks=true,       
    linkcolor=red,          
    citecolor=darkgreen,        
    filecolor=magenta,      
    urlcolor=linkblue,           
    menucolor=blue,
}

\newcommand\T{\rule{0pt}{2.6ex}} 

\begin{document}
\title{\TITLE}
\newcommand{\sunysb}{Department of Chemistry, Stony Brook University, Stony Brook, NY 11794, USA}
\newcommand{\bnl}{Physics Department, Brookhaven National Laboratory, Upton, NY 11796, USA}
\newcommand{\ct}{Department of Physics, University of Cape Town, Private Bag X3, Rondebosch 7701, South
Africa}
\author{Jiangyong Jia}\email[Correspond to ]{jjia@bnl.gov}\affiliation{\sunysb}\affiliation{\bnl}
$\phantom{\email{}}$
\author{W. A. Horowitz}\email{wa.horowitz@uct.ac.za}\affiliation{\ct}
\author{Jinfeng Liao}\email{jliao@bnl.gov}\affiliation{\bnl}
\date{\today}
\begin{abstract}
Focusing on four types of correlation plots, $R_{\rm AA}$ vs.\
$v_2$, $R_{\rm AA}$ vs.\ $I_{\rm AA}$, $I_{\rm AA}$ vs.\
$v_2^{I_{\rm AA}}$ and $v_2$ vs.\ $v_2^{I_{\rm AA}}$, we
demonstrate how the centrality dependence of \emph{correlations} between multiple jet quenching
observables provide valuable insight into the energy loss mechanism
in a quark-gluon plasma.  In particular we find that a qualitative energy loss model gives a
good description of $R_{\rm AA}$ vs.\ $v_2$ only when we take $\Delta
E\sim l^3$ and a medium geometry generated by a model of the Color Glass
Condensate. This same $\Delta E\sim l^3$ model also qualitatively
describes the trigger $p_T$ dependence of $R_{\rm AA}$ vs.\ $I_{\rm
AA}$ data and makes novel predictions for the centrality dependence
for this $R_{\rm AA}$ vs.\ $I_{\rm AA}$ correlation. Current data
suggests, albeit with extremely large uncertainty, that
$v_2^{I_{\rm AA}}\gg v_2$, a correlation that is difficult to
reproduce in current energy loss models.
\end{abstract}
 \maketitle

{\bf Introduction:} The combination of theoretical predictions and experimental measurements of \highpt jet observables provides a unique basis for determining the properties of the strongly-interacting quark
gluon plasma (sQGP) created in Au+Au collisions at the Relativistic
Heavy Ion Collider (RHIC) \cite{Gyulassy:2004zy,Wiedemann:2009sh,Majumder:2010qh}.  After nearly a
decade long effort, jet quenching via final state partonic interactions, as an experimental phenomenon,
has been firmly established at RHIC~\cite{Majumder:2010qh}. The challenge for theory is to find an energy loss model built on first principles derivations that \emph{simultaneously} describes the known observables. Currently no such model exists, and there is a debate over the exact, or even dominant, energy loss mechanism at work in the sQGP (see, e.g.,~\cite{Majumder:2010qh,Horowitz:2010yi,Jia:2010hg}).

Four observables of interest in leading particle quenching physics are single hadron suppression $R_{\rm AA}$, $v_2$ (one half the coefficient of the $\cos(2\phi_s)$ term in the Fourier expansion of the suppression relative to the reaction plane (RP) $R_{\rm AA}(\phi_s=\phi-\Psi_{\rm RP})$), di-hadron suppression $I_{\rm AA}$, and $v_2^{I_{\rm AA}}$. These observables are interesting because they probe the same energy loss processes in a heavy ion collision, but with different underlying parton spectra and/or path length ``$l$'' dependencies. For example $R_{\rm AA}$ at different $\phi_s$ has identical input parton spectra but explores different path lengths; on the other hand, $I_{\rm AA}$ probes a harder input spectrum and a different set of paths compared to $R_{\rm AA}$. The importance of using multiple observables to constrain the possible energy loss mechanism in heavy ion collisions is well known~\cite{Zhang:2007ja,Dominguez:2008vd,Bass:2008rv,Renk:2008xq,Renk:2011wp}.  However, other than one previous publication which examined the centrality dependence of  $R_{\rm AA}$ vs $v_2$~\cite{Horowitz:2005ja}, the comparison between theory and data was made one observable at a time and usually as a function of \pt for one centrality selection.  

In this work we argue that correlating these observables directly against each other and studying the centrality dependence of the correlation provides novel insights into the \highpt energy loss mechanism. We propose four types of correlations that can be studied experimentally at RHIC and the LHC: $R_{\rm AA}$ vs.\ $v_2$, $R_{\rm AA}$ vs.\ $I_{\rm AA}$, $I_{\rm AA}$ vs.\ $v_2^{I_{\rm AA}}$ and $v_{2}$ vs.\ $v_2^{I_{\rm AA}}$. We shall first give a brief overview of the experimental measurements of each observable at RHIC. We then explore the main features of the four correlations revealed from the experimental data. Using a jet absorption model, we demonstrate the importance of these correlations by exploring their sensitivities on the geometry and parton spectra shape. We conclude with a discussion of how these correlations can be used to disentangle the ``$l$'' dependence of energy loss from the collision geometry and parton spectra.


{\bf Overview of experimental results and theoretical comparisons:}
The most precise measurements of high-$p_T$ single hadron suppression and anisotropy
were carried out by the PHENIX experiment using
$\pi^0$ mesons \cite{Adare:2008qa,Adare:2010sp}, reaching $p_T\sim 20$
GeV/$c$ for $R_{\rm AA}$ and beyond 10 GeV/$c$ for $v_2$. $R_{AA}$ is defined as
\be
R_{AA}^h(p_T,\,\phi_s,\,b)\equiv\frac{\frac{dN^{AA\rightarrow h+X}}{d^2p_T(p_T,\,\phi_s,\,b)}}{N_{bin}(b)\,\frac{dN^{pp\rightarrow h+X}}{d^2p_T(p_T)}},
\ee
where $N_{bin}(b)$ is the number of binary (hard scattering $pp$-like) collisions at impact parameter $b$, and $v_2\equiv\int d\phi_s \, R_{\rm AA}(\phi_s)\cos(2\phi_s)/\int d\phi_s \, R_{\rm AA}(\phi_s)$.  The
$R_{\rm AA}$ shows an almost $p_T$-independent factor of 5
suppression in central collisions for $p_T>4$ GeV/$c$. The $v_2$
drops from 3 to 7 GeV/$c$, but remains positive at higher $p_T$.
Current jet quenching models based on the pQCD framework, when
tuned to $R_{\rm AA}$ data, significantly under-predict the
$v_2$~\cite{Adare:2010sp}. In contrast, non-perturbative
approaches, for example those based on AdS/CFT gauge gravity
duality~\cite{Marquet:2009eq}, seem to work well. The data seem to
prefer the $\Delta E\sim l^3$ path length dependence, a result based on AdS/CFT
~\cite{Dominguez:2008vd}, as opposed to the quadratic dependence
$\Delta E\sim l^2$ predicted radiative energy
loss predicted by pQCD~\cite{Gyulassy:1993hr}. Alternatively, a simultaneous
description of $R_{AA}$ and $v_2$ may also be achieved via a
late-stage non-perturbative effect near the QCD confinement
transition~\cite{Horowitz:2005ja,Liao:2008dk}.

\begin{figure}[ht]
\centering
\epsfig{file=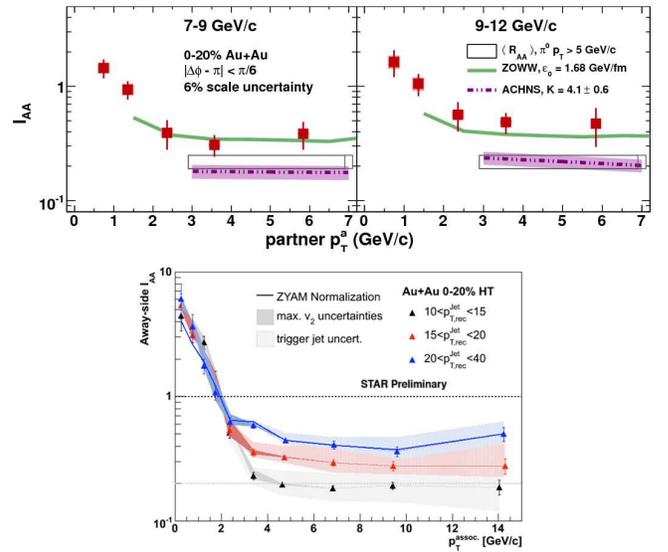,width=1.0\linewidth}
\caption{\label{fig:2} (Color online) (Top panels) PHENIX $\pi^0-h$ correlation results with $7<p_T^{t}<12$ GeV/$c$
and $0.5<p_T^a<7$ GeV/$c$~\cite{Adare:2010ry}.
(Bottom panel) the STAR jet-h correlation results with reconstructed trigger jet momentum in $10<p_{T,rec}^{jet}<40$ GeV/$c$~\cite{Putschke}.
Both are for the 0-20\% Au+Au centrality bin. }
\end{figure}

\begin{figure}[ht]
\centering
\epsfig{file=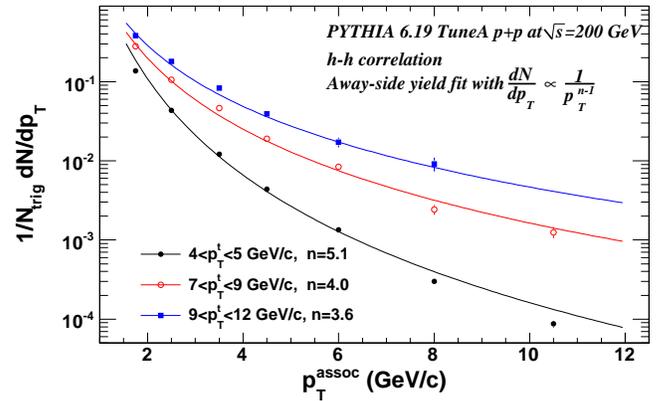,width=1.0\linewidth}
\caption{\label{fig:1b} (Color online) PYTHIA simulation of the away-side per-trigger charged hadron yield spectra in PHENIX $\eta$ acceptance
for three different charged hadron trigger $p_T$ ranges. The away-side yields are parameterized with a power law function.}
\end{figure}

The suppression of the away-side jet is quantified by $I_{\rm AA}$,
the ratio of the per-trigger yield (away-side jet multiplicity
normalized by number of triggers) in Au+Au collisions to that in
p+p collisions. Pure geometrical considerations would imply $I_{\rm
AA}<R_{\rm AA}$, due to a longer path length traversed by the
away-side jet. But recent PHENIX~\cite{Adare:2010ry} and
STAR~\cite{Putschke} measurements show that $I_{\rm AA}$ is
\emph{constant} for associated hadron $p_T^a>3$ GeV/$c$, within the current experimental uncertainties, and this
constant level is above the $R_{\rm AA}$ for the trigger hadrons,
i.e.\ $I_{\rm AA}>R_{\rm AA}$ (see Fig.~\ref{fig:2}). Furthermore,
the constant level of $I_{\rm AA}$ increases for higher trigger
$p_T^t$. This result can be qualitatively explained by the bias of
the away-side jet energy by the trigger $p_T$: as we show with a PYTHIA simulation in \fig{fig:1b}, the initial
away-side jet spectra becomes harder as higher $p_T$ triggers are
required.  As we discuss in more detail below, the harder the input spectrum, the larger the fractional energy loss 
is required for the same $I_{\rm AA}$ value. The ACHNS
model \cite{Armesto:2009zi} results, constrained by the $R_{\rm
AA}$ data, are incompatible with the $I_{\rm AA}$ values shown in
\fig{fig:2}; while the ZOWW model~\cite{Zhang:2009rn}
seems to describe the $I_{\rm AA}$ data alone shown in
\fig{fig:2}, it too fails at simultaneously
describing both $R_{\rm AA}$ and $I_{\rm AA}$~\cite{Nagle:2009wr}.

PHENIX~\cite{Adare:2010mq} recently reported the first measurement of the anisotropy of away-side suppression,
$v_2^{I_{\rm AA}} \equiv \int d\phi_s \, I_{\rm AA}(\phi_s)\cos(2\phi_s)/\int d\phi_s \, I_{\rm AA}(\phi_s)$. The away-side yield shows a strong
variation with angle of the trigger relative to the RP. This
variation is much larger than that for inclusive $\pi^0$ in the
same trigger $p_T$ range. The current measurement is statistics
limited; however the result is tantalizing as energy loss models
usually predict much smaller $v_2^{I_{\rm
AA}}$~\cite{Adare:2010mq}.

{\bf The jet absorption model:} We use a model from~\cite{Drees:2003zh,Jia:2010ee} to investigate correlations
between the four observables, and to check the sensitivities of
these correlations to the collision geometry and $l$ dependence of
the energy loss. The model is based on a na\"ive jet absorption
picture where the fractional energy loss of a high $p_T$ particle
is proportional to a line integral $I$ through the medium,
$\epsilon = \tilde{\kappa} I$, where $\epsilon = 1 - p_T^f/p_T^i$. For a power law production spectrum with index of $n$,
$dN/d^2p_T\sim p_T^{-n}$, $R_{\rm AA}$ is related to the fractional
energy loss $\epsilon=\Delta E/E$
via~\cite{Adcox:2004mh,Horowitz:2010dm}:
\begin{eqnarray}
R_{\rm AA} = \langle(1-\epsilon)^{n-2}\rangle \approx \langle e^{-(n-2)\epsilon}\rangle \approx \langle e^{-\kappa I}\rangle,
\label{eq:1}
\end{eqnarray}
where $\kappa = (n-2)\tilde{\kappa}$, and $\langle\cdots\rangle$
indicates an average over the binary collision profile.  The line integral $I$ is calculated as $I_1=\int \rho
\hspace{1mm}dl$ or $I_2=\int \rho \hspace{1mm}ldl$. The former
corresponds to a quadratic dependence of energy loss ($dE\sim ldl$)
in a longitudinally expanding medium
($\rho(\tau)\sim1/\tau\sim1/l$), while the latter corresponds to a
cubic dependence ($dE\sim l^2dl$) of energy loss in a longitudinally
expanding medium. Up to slowly varying logarithmic factors the interference of the unmodified vacuum radiation associated with the production of a \highpt parton with the medium-induced bremsstrahlung radiation in the deep Landau-Pomeranchuk-Migdal region---which one expects with the ordering of length scales $1/\mu\ll\lambda_{mfp}\ll l$ in a weakly-coupled QGP described by Hard Thermal Loop pQCD weakling interacting with the \highpt parent parton that we might expect in Au+Au collisions at RHIC---yields a fractional energy loss that scales with the square of the pathlength, $d\epsilon\propto ldl$.  For the not-too-large fractional energy losses at RHIC, $\langle\epsilon\rangle\sim0.2$ for $R_{AA}\sim0.2$, the exponential absorption model is a reasonable approximation to the $1-\epsilon$ Jacobian expect for pQCD energy loss.  On the other hand, under the assumption that all the couplings between the sQGP and the \highpt parton are very large and the dominant physics can be well approximated using the AdS/CFT conjecture, then the thermalization distance for a light \highpt probe parton scales as $E^{1/3}$.  The exponential model can again be used, in this case capturing the physics of the probability of escape for the strongly coupled \highpt particles.  
%
%
We model the medium density $\rho$ either by the
participant density profile from Glauber geometry or gluon density
profile from CGC geometry~\cite{Drescher:2006pi}. The dominant effect of
event-by-event fluctuations in the sQGP are included in a medium rotation
procedure~\cite{Alver:2008zza}.  Comparison of the jet absorption model to data is a reasonable first step when examining the physics of the centrality dependence of the correlations investigated in this paper as it captures both the pathlength dependence and medium geometry effects.  

$\kappa$ is the only free parameter in our energy loss model; we tune
it to reproduce $R_{\rm AA}\sim0.18$ for 0-5\% most central
$\pi^0$.  Once $\kappa$ is fixed, we then predict the
centrality dependence of $R_{\rm AA}$ in 5\% centrality increments, as well as
$I_{\rm AA}$, $v_2$, and $v_2^{I_{\rm AA}}$. The
$\kappa$ values for the four cases (the combinations of $l^2$,
$l^3$ and Glauber, CGC media) are summarized in Table~\ref{tab:1}. Note
that for a given ``$l$'' dependence, the suppression level is
essentially controlled by the product of $\kappa$ and the average
matter density $\langle\rho_{\rm
medium}\rangle\equiv\int\rho(\vec{\mathbf{x}})^2 d^2x/\int
\rho(\vec{\mathbf{x}}) d^2x$ in the 0-5\% centrality bin. In
general the $\kappa\langle\rho_{\rm medium}\rangle$ for CGC
geometry is slightly larger than Glauber geometry, primarily
because the former has a smaller matter profile~\cite{Jia:2010ee},
while both geometries are assumed to have the same binary collision
profile.

\begin{table}[ht]
\caption{\label{tab:1} $\kappa$, average matter density
$\langle\rho_{\rm medium}\rangle=\int\rho(\vec{\mathbf{x}})^2
d^2x/$ $\int \rho(\vec{\mathbf{x}}) d^2x$%
, and the product of the
two in the 0\%-5\% Au+Au centrality bin, for the four cases calculated
in our study. }
\begin{tabular}{l|c|c|c}
\hline
& $\quad\kappa\quad$ & $\;\;\langle\rho_{\rm medium}\rangle\;\;$ & $\;\;\kappa\langle\rho_{\rm medium}\rangle\;\;$ \\
\hline
$l^2$  Glauber  \T &0.147   & 2.96                 & 0.452\\
$l^2$  CGC      \T &0.076   & 6.40                 & 0.486\\
$l^3$  Glauber  \T &0.082   & 2.96                 & 0.243\\
$l^3$  CGC      \T &0.046   & 6.40                 & 0.294\\\hline
\end{tabular}
\end{table}

If we assume that the di-hadron production spectrum is also a power
law, $dN/(d^2p_T^ad^2p_T^t)\sim (p_T^a)^{-n_a}(p_T^t)^{-n_t}$ then
\begin{align}
I_{\rm AA} & = \langle(1-\epsilon_a)^{n_a-2}(1-\epsilon_t)^{n_t-2}\rangle/R_{\rm AA} \nonumber\\
& \approx \langle e^{-\kappa_{\rm away}I_a}e^{-\kappa I_t}\rangle/R_{\rm AA}.
\label{eq:I1}
\end{align}
Since $\kappa\propto n_t-2 = n-2$ according to Eq.~\ref{eq:1}, the
effective $\kappa_{\rm away}$ for away-side jets should be smaller
due to a smaller $n_a$. \fig{fig:1b} shows that increasing the momentum of the trigger particle from $4-5$ GeV/c to $9-12$ GeV/c yields a reduction in the power law for the away-side spectrum from $n_a = 5.1$ to $n_a=3.6$.  One may effectively model this reduction in away-side input spectrum in our absorption model by approximating
\be
\label{eq:kapaway}
\kappa_{\rm away}=\frac{n_a-2}{n-2}\,\kappa;
\ee
hence the effective strength of the energy loss is reduced by a factor between 2 and 4 for the trigger particle momentum ranges currently measured. One may readily see that as the trigger momentum range is increased, the effects of energy loss on the suppression of particles is reduced; $I_{AA}$ is not simply smaller than $R_{AA}$ due to the longer pathlengths that result from the trigger bias in the di-hadron measurement.  

\begin{figure}[ht]
\centering \epsfig{file=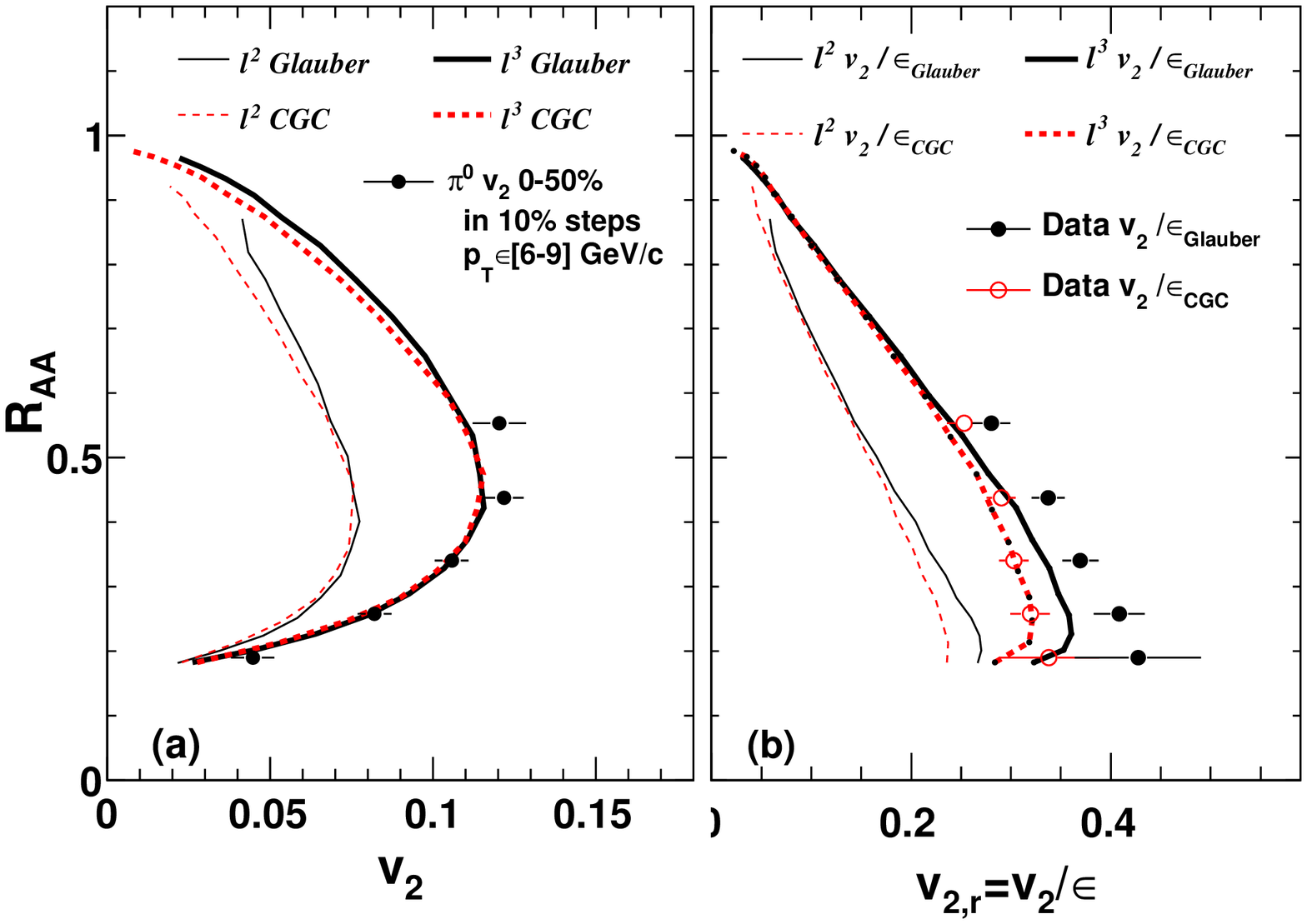,width=1\linewidth}
\epsfig{file=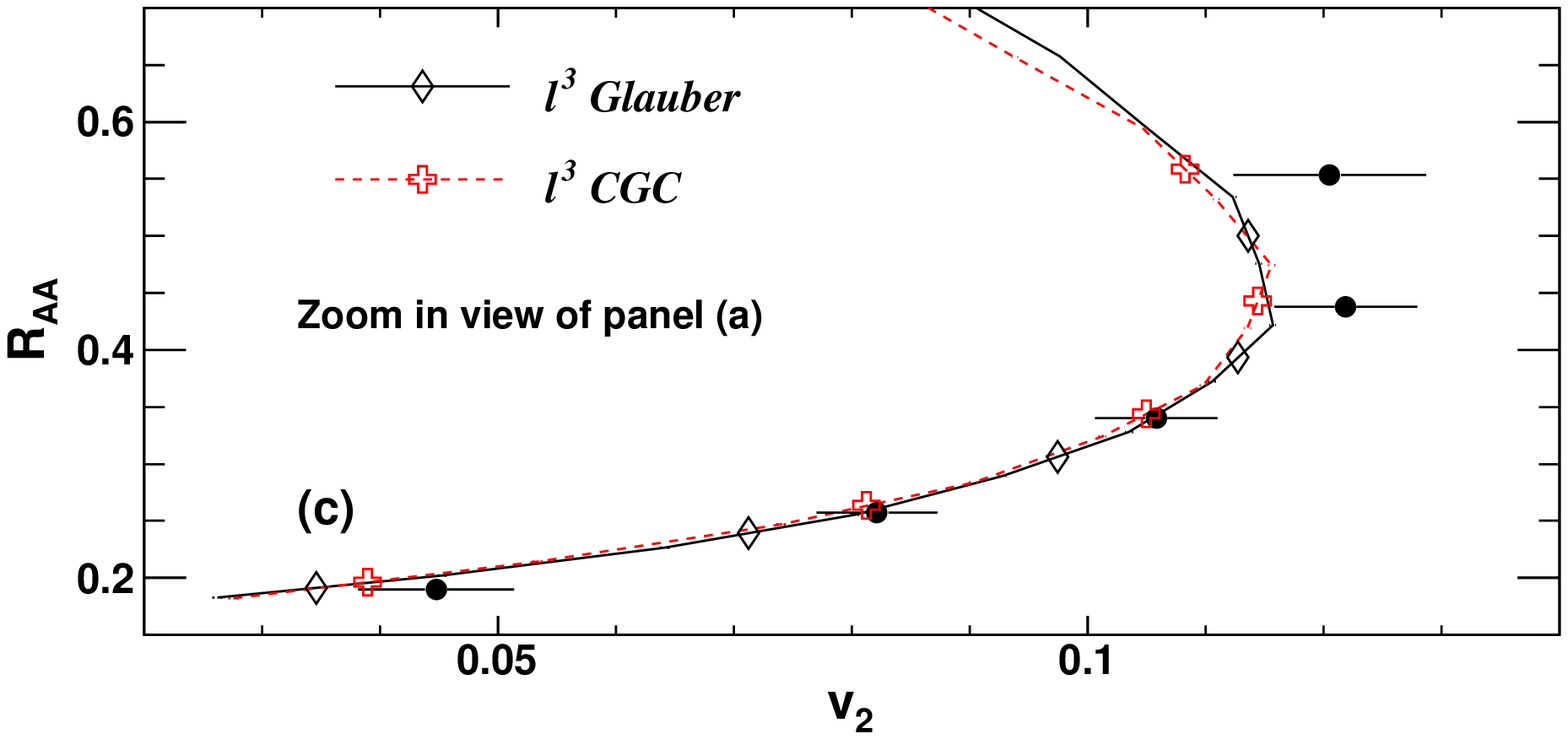,width=1\linewidth}
\caption{\label{fig:3} (Color online)
(a) Centrality dependence of $R_{\rm AA}$ vs.\ $v_2$ from the jet absorption model in
5\% centrality steps and data (solid circles); only statistical errors are shown. (b) The same data and calculations,
except $v_2$ is divided by the eccentricity. (c) The same as in (a) but zoomed in and with centrality binning explicitly shown for the theoretical results.}
\end{figure}

{\bf Results:} \fig{fig:3} (a) shows the predicted
correlation between $R_{\rm AA}$ vs.\ $v_2$ from the jet absorption
model over the full centrality range. The calculations appear to
show little dependence on the assumed geometry (more later), but
clearly $v_2$ increases dramatically from $l^2$ to $l^3$
dependence. The $l^3$ dependence agrees with data well, implying
that it can simultaneously describe both $R_{\rm AA}$ and $v_2$, a
conclusion already made in~\cite{Jia:2010ee}.

We know that the low-$p_T$ $v_2$ is observed to scale with
eccentricity ($\epsilon$)~\cite{Adare:2010ux}. It was shown
in~\cite{Jia:2010ee} that the jet quenching $v_2$ also
approximately scales with $\epsilon$. Thus it is instructive to
plot $R_{\rm AA}$ versus the reduced quantity $v_{2,r}=
v_2/\epsilon$, as shown in \fig{fig:3} (b). The data now
appear as two sets of points, filled black circles and open red circles, corresponding to Glauber geometry or
CGC geometry, respectively. They both indicate an anti-correlation
with $R_{\rm AA}$, that is a large $v_{2,r}$ corresponds to a small
$R_{\rm AA}$ and vice versa.  Similar trends are also shown by the
calculations: as quenching becomes stronger the
surviving jets further amplify the initial asymmetry. In \fig{fig:3} (c) we show the centrality binned theoretical results as open black diamonds and open red crosses for the $l^3$ Glauber and CGC medium results respectively; these theory points should be directly compared to the filled black circle data points.  

Note that while the $\epsilon\sim l^3$ models with either a CGC
medium or Glauber medium appear to describe the $R_{\rm AA}$ vs.\
$v_2$ data in \fig{fig:3} (a) well, only the cubic model with
the CGC medium describes the $R_{\rm AA}$ vs.\ $v_{2,r}$ data shown
in \fig{fig:3} (b).  As shown in \fig{fig:3} (c), this
is because the $l^3$ model with Glauber medium does not describe
the data at the \emph{correct centrality bin} whereas the $l^3$
model with CGC medium does. Eccentricity is a centrality dependent
quantity (as discussed in detail in \cite{Jia:2010ee} the CGC
geometry is smaller relative to the Glauber geometry), and
normalizing with respect to eccentricity has emphasized this
mismatch in the centrality-binned results for theory and data. Thus $R_{\rm AA}$
versus $v_{2}/\epsilon$ can better illustrate the relation between
jet quenching and azimuthal anisotropy, indicating here that our
model only describes the $R_{\rm AA}$ vs.\ $v_2$ data for
AdS/CFT-like energy loss in a CGC medium.

\begin{figure}[ht]
\centering \epsfig{file=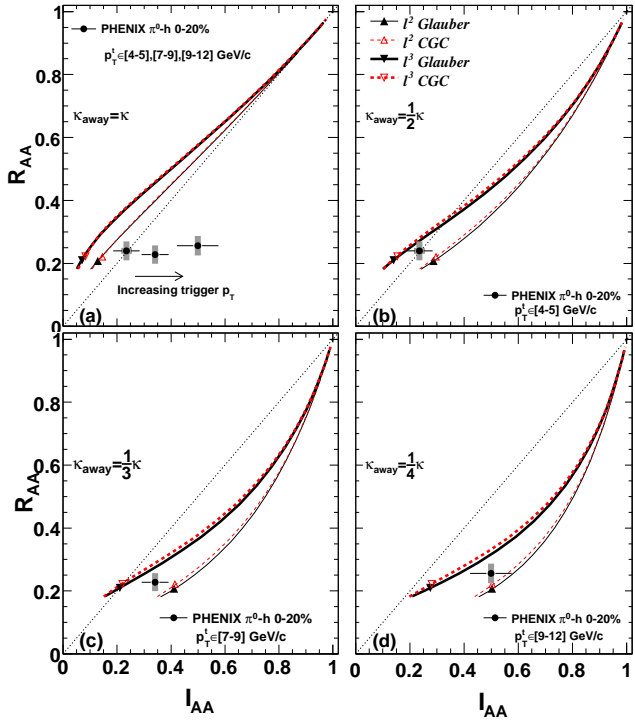,width=1.0\linewidth}
\caption{\label{fig:4} (Color online) Centrality dependence of $R_{\rm AA}$ vs.\ $I_{\rm AA}$ from the jet absorption model,
compared with PHENIX data~\cite{Adare:2010ry} in the 0-20\% centrality bin.
The calculations were done assuming $\kappa$ for away-side jet is (a) the same, (b) 1/2, (c) 1/3, or (d) 1/4 of that for inclusive jets.
We compare the model curves in (b) to the left-most ([4-5] GeV/c trigger), (c) to the central ([7-9] GeV/c trigger),
and (d) to the right-most ([9-12] GeV/c trigger) PHENIX data point. The dotted diagonal line indicates $I_{\rm AA}=R_{\rm
AA}$. The explicit predictions for 0-20\% centrality are shown as diamond
and cross symbols.}
\end{figure}

\fig{fig:4} shows the correlation between $R_{\rm AA}$ and
$I_{\rm AA}$ from the jet absorption model, calculated over the
full centrality range. The model results are compared with STAR and
PHENIX data from \fig{fig:2} (0-20\% centrality bin) with
their $I_{\rm AA}$ values integrated over $p_T^{a}>3$ GeV/$c$
(where $I_{\rm AA}$ is flat). In \fig{fig:4} (a), the data points have roughly the same
$R_{\rm AA}$ value, but are spread out in $I_{\rm AA}$ for
different trigger momenta, $p_T^t$. The reason, as explained in the
discussion of \fig{fig:2} \& \fig{fig:1b}, is that $I_{\rm
AA}$ depends not only on the path length but also on the shape of
the input spectra, and the larger the trigger momentum $p_T^t$ the
harder the away-side spectrum. \fig{fig:4} (a) also shows that $I_{\rm AA}<R_{\rm AA}$ when only
the path length effect is included (i.e.\ we take $\kappa_{\rm
away}=\kappa$).  We then attempt to model the effect of the trigger
bias on the hardening of the away-side spectrum
(see \fig{fig:1b}) by using \eq{eq:kapaway}, which yields $\kappa_{\rm away} = \kappa/2$ in
\fig{fig:4} (b) ($p_T^t\in[4-5]$ GeV/c), $\kappa_{\rm away} = \kappa/3$ in \fig{fig:4} (c) ($p_T^t\in[7-9]$ GeV/c), and $\kappa_{\rm away}
= \kappa/4$ in \fig{fig:4} (d) ($p_T^t\in[9-12]$). The toy model improves its agreement with the
PHENIX data when both the path length and spectral dependencies are
included. 

In particular the $l^3$ AdS/CFT-like energy loss model
that described the $R_{\rm AA}$ vs.\ $v_2$ data so well appears to
describe the $R_{\rm AA}$ vs.\ $I_{\rm AA}$ data to within about 2-3
standard deviations.  One also again sees that the CGC medium
yields results whose centrality dependence is in slightly better agreement
with the data than the results from the Glauber medium. However it
is clear that the $l^3$ models systematically under-predict the
$I_{\rm AA}$ data.  On the other hand the $l^2$ models tend to
disagree more on the level of 1 standard deviation.  That the $l^2$
models tend to describe the normalization and correlation--but not
anisotropy--of $R_{\rm AA}$ and $I_{\rm AA}$ might suggest the
importance of hadronization or flow-coupling effects that are
neglected in our model~\cite{Horowitz:2005ja,Liao:2008dk,Horowitz:2011}.

In general the toy model predicts significantly different $R_{\rm
AA}$ vs.\ $I_{\rm AA}$ curves as a function of centrality as a
function of trigger momentum: the larger $p_T^t$ (or, equivalently, smaller
$\kappa_{\rm away}$), the more concave the $R_{\rm AA}$ vs.\ $I_{\rm
AA}$ curve becomes. 
The concavity of the correlation is also in general larger for a CGC medium than for a Glauber medium.  It would be interesting to see these predictions compared with future measurements performed over the full
centrality range in small centrality bins.

\begin{figure}[ht]
\centering \epsfig{file=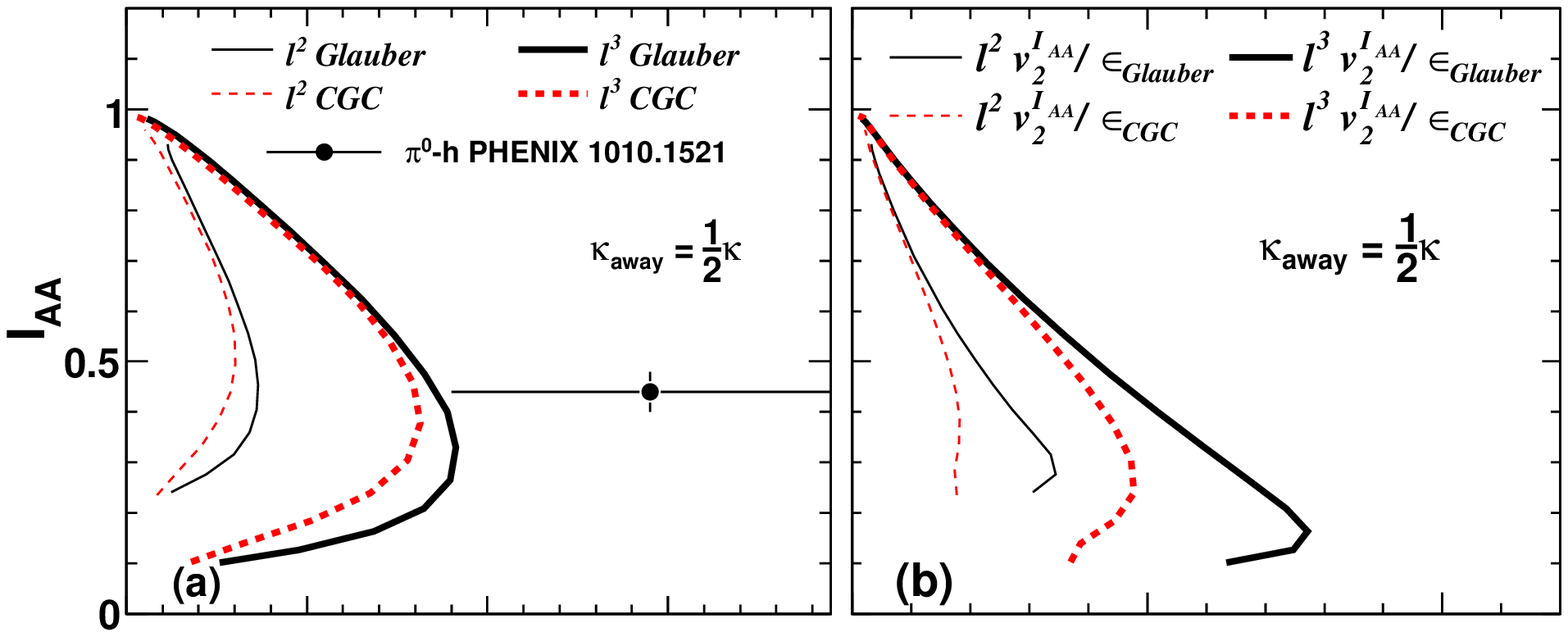,width=1\linewidth}\vspace*{-0.65cm}
\centering \epsfig{file=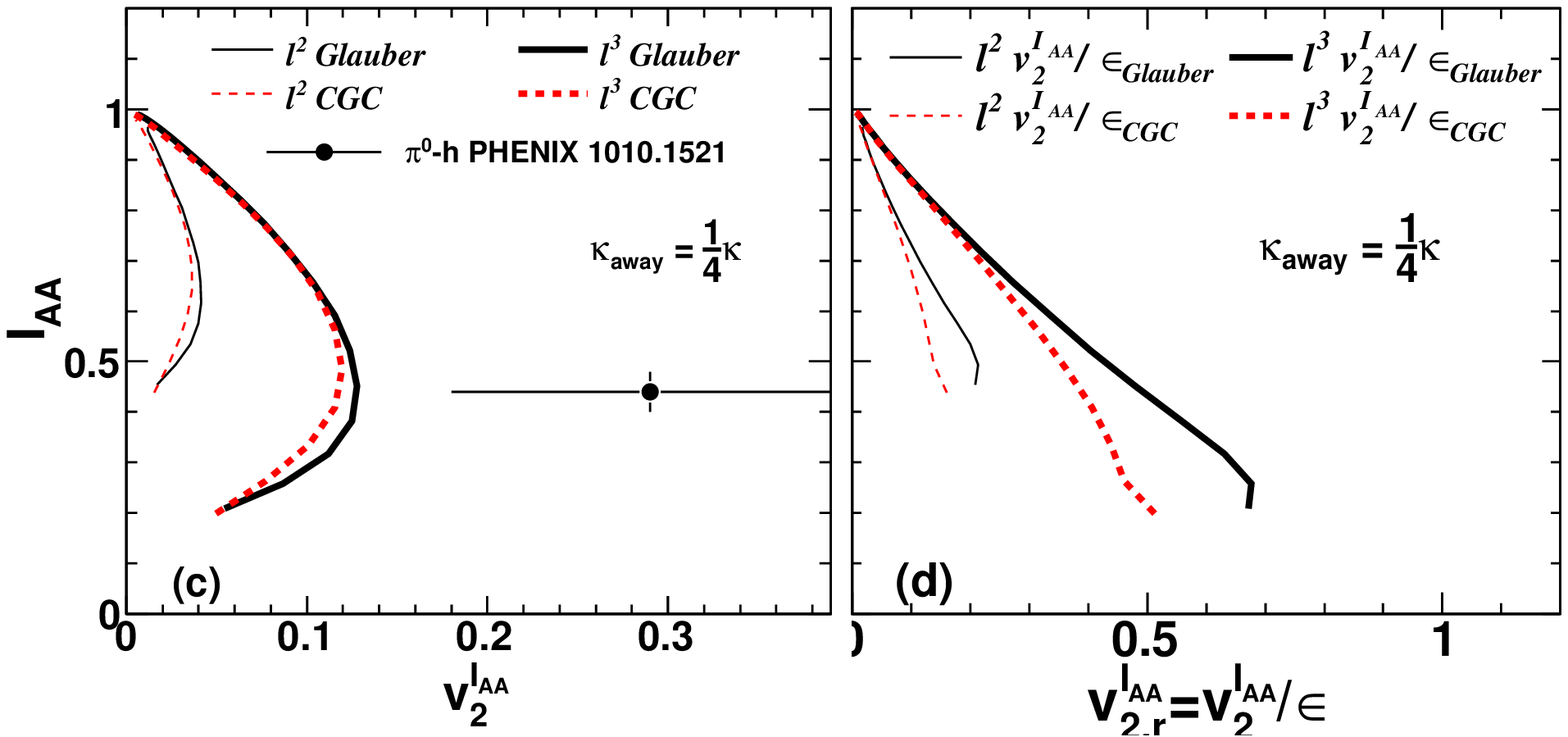,width=1\linewidth}
\caption{\label{fig:5} (Color online) (a) Centrality dependence of $I_{\rm AA}$ vs.\ $v_2^{I_{\rm AA}}$ from jet absorption model in 5\%
centrality steps and data (solid circles).
(b) The same data and calculations,
excepts their $v_2^{I_{\rm AA}}$ have been divided by the eccentricities. (c), (d) are same as (a),
(b) but are calculated for $\kappa_{\rm away}=\frac{1}{4}\kappa$.}
\end{figure}

The strong suppression of the away-side jet should also lead to an
anisotropy of the $I_{\rm AA}$ relative to the RP. 
\fig{fig:5} (a) shows the predicted
correlation between $I_{\rm AA}$ and $v_2^{I_{\rm AA}}$, assuming
$\kappa_{\rm away}=\frac{1}{2}\kappa$. The corresponding correlation
between $I_{\rm AA}$ and reduced anisotropy $v_{2,r}^{I_{\rm
AA}}=v_2^{I_{\rm AA}}/\epsilon$ is shown in \fig{fig:5} (b).
The results for $\kappa_{\rm away}=\frac{1}{4}\kappa$ are shown in
\fig{fig:5} (c)-(d).

Several interesting features can be identified from the figure. First, $v_2^{I_{\rm AA}}$ increases dramatically from $l^2$ to
$l^3$ dependence, but even the $l^3$ model under-predicts the
PHENIX data~\cite{Adare:2010mq} by about one standard deviation, where the uncertainty is currently very large. Second,
calculated $v_2^{I_{\rm AA}}$ values are very sensitive to
$\kappa_{\rm away}$: they are larger than the inclusive hadron
$v_2$ of \fig{fig:3} for $\kappa_{\rm away}=\frac{1}{2}\kappa$,
but are less for $\kappa_{\rm away}=\frac{1}{4}\kappa$. It will therefore be very useful to see experimental results for $v_2^{I_{\rm AA}}$ as a function of trigger momentum.  Finally, we
also see a strong anti-correlation between $v_{2}^{I_{\rm
AA}}/\epsilon$ and $I_{\rm AA}$, quite similar to that between
$v_{2}/\epsilon$ and $R_{\rm AA}$. Such similarity may not be
surprising if the physics of jet quenching for the trigger and away
jets were identical (as in the present model calculation). A
precision measurement of these two correlations, therefore, could
either confirm such similarity or suggest new physics in the
away-side jet quenching.

The correlation between $I_{\rm AA}$ and $v_2^{I_{\rm AA}}$ is also
quite sensitive to the choice of the collision geometry: switching
from Glauber geometry to CGC geometry leads to about a 20\%
reduction of $v_{2}^{I_{\rm AA}}$ and $v_{2}^{I_{\rm AA}}/\epsilon$
at fixed $I_{\rm AA}$ for $\kappa_{\rm away}=\frac{1}{2}\kappa$
(significantly smaller for $\kappa_{\rm away}=\frac{1}{4}\kappa$). Thus
a precise measurement of this correlation may help in
distinguishing between different initial geometries.

\begin{figure}[ht]
\begin{center}
\epsfig{file=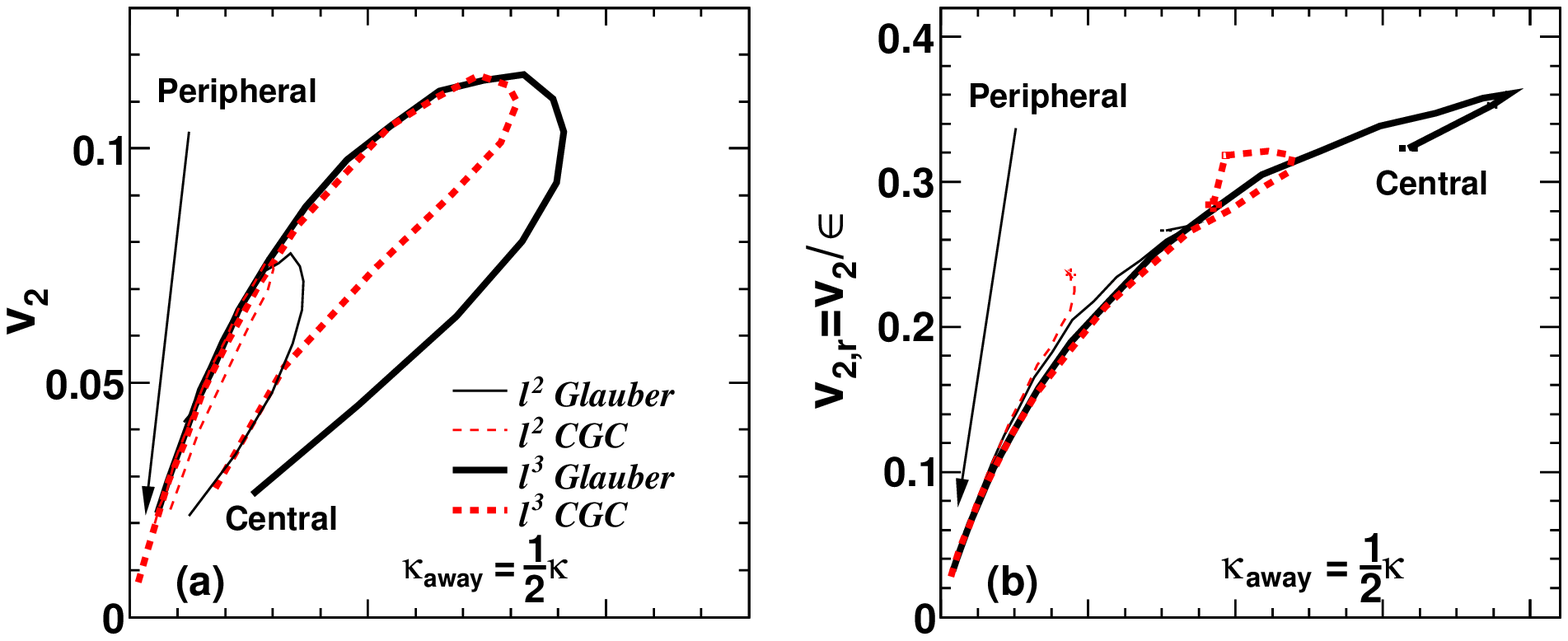,width=1\linewidth}\vspace*{-0.65cm}
\epsfig{file=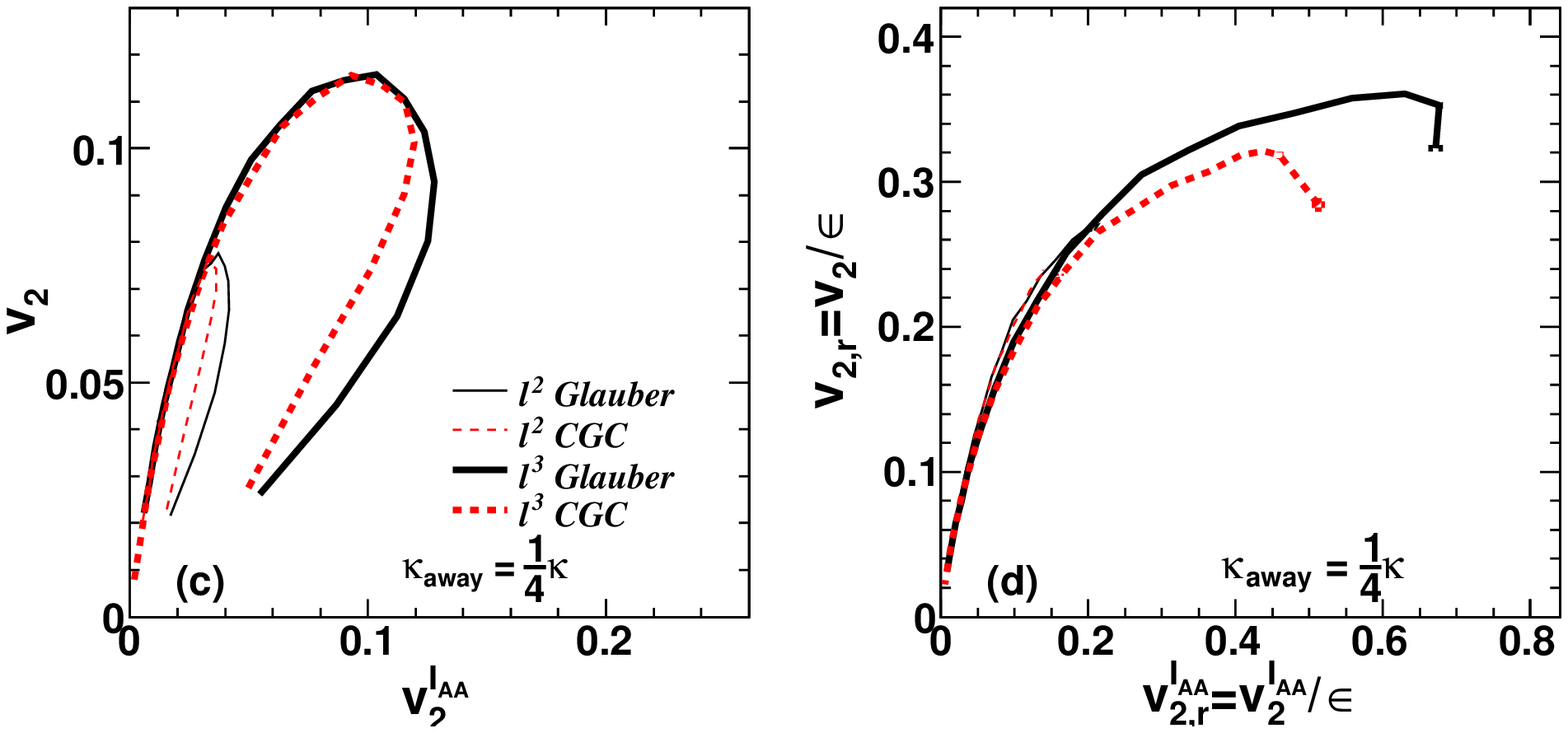,width=1\linewidth}
\end{center}
\caption{\label{fig:6} (Color online) (a) $v_2$ vs.\ $v_2^{I_{\rm
AA}}$ and (b) $v_{2,r}$ vs.\ $v_{2,r}^{I_{\rm AA}}$ for 5\%
centrality steps and four cases for
$\kappa_{\rm away}=\frac{1}{2}\kappa$. (c), (d) are same as (a), (b)
but are calculated for $\kappa_{\rm away}=\frac{1}{4}\kappa$.
The scaling seen in (b) and (d) is broken in more central collisions
as the fluctuations in eccentricity $\epsilon$ dominate the mean, $\langle\epsilon\rangle\ll
\sqrt{\left(\epsilon-\langle\epsilon\rangle\right)^2}$.
}
\end{figure}

\fig{fig:6} shows the correlation between $v_2$ and
$v_2^{I_{\rm AA}}$ for $\kappa_{\rm away}=\frac{1}{2}\kappa$ and
$\kappa_{\rm away}=\frac{1}{4}\kappa$. Since both observables first
increase then decrease from peripheral to central collisions, the
correlation plot shows a rather sharp turn at around 20-30\% centrality
($N_{\rm part}\sim150$). This provides a rather precise way of
identifying the centrality range at which the anisotropy reaches a
maximum. The right panel shows the correlation of reduced
anisotropies. We see that all four scenarios fall approximately on
a universal curve, especially for
$\kappa_{\rm away}=\frac{1}{2}\kappa$, but with different reaches along
the curve. Apparently, the reach is larger for Glauber geometry and
higher order $l$ dependence. This implies that the efficiency with
which jet quenching converts the eccentricity into anisotropy
depends on collision geometry and $l$ dependence. This curve also
has an interesting shape: At small $v_{2,r}$ ($\lesssim 0.2$ for
$\kappa_{\rm away}=\frac{1}{2}\kappa$ and $\lesssim 0.3$ for
$\kappa_{\rm away}=\frac{1}{4}\kappa$), which corresponds to more
peripheral collisions, $v_{2}$ is larger than $v_{2}^{I_{\rm AA}}$;
but then $v_{2}^{I_{\rm AA}}>v_{2}$ at large $v_{2,r}$ ($>0.2$ for
$\kappa_{\rm away}=\frac{1}{2}\kappa$ or $>0.3$
 for $\kappa_{\rm away}=\frac{1}{4}\kappa$)~\footnote{However for $l^2$ and
CGC geometry, $v_{2}>v_{2}^{I_{\rm AA}}$ in all centrality bins}.

Before closing this section, we want to discuss all correlations
together. In general, jet quenching models predict an
anti-correlation between suppression and anisotropy ($R_{\rm AA}$
vs.\ $v_2$, and $I_{\rm AA}$ vs.\ $v_2^{I_{\rm AA}}$), while they
predict a correlation between the suppressions ($R_{\rm AA}$ vs.\
$I_{\rm AA}$) and between the anisotropies ($v_2$ vs.\ $v_2^{I_{\rm
AA}}$). Thus it is rather surprising to see from the data that
$I_{\rm AA}$ has a larger anisotropy than that for $R_{\rm AA}$ ($v_2^{I_{\rm AA}}>v_2$), yet it is less suppressed.
This feature is very difficult to reproduce in our simple model (see \fig{fig:4} (d) and
\fig{fig:5} (c)), and, in general, is a challenge for jet quenching
theory~\cite{Nagle:2009wr}.

{\bf Summary:} The simultaneous description of multiple observables
tightly constrains jet quenching models. In this paper we identified four correlations among 
single hadron and di-hadron observables as useful for constraining
jet quenching models: $R_{\rm AA}$ vs.\ $v_2$, $R_{\rm AA}$ vs.\
$I_{\rm AA}$, $I_{\rm AA}$ vs.\ $v_2^{I_{\rm AA}}$ and $v_2$ vs.\
$v_2^{I_{\rm AA}}$. Using a jet absorption model, we showed that
these correlations are sensitive to various ingredients in the jet
quenching calculations, such as the path length dependence,
collision geometry, and input spectra shape. Specifically, $R_{\rm
AA}$ vs.\ $v_2$ is most sensitive to $l$ dependence, $R_{\rm AA}$
vs.\ $I_{\rm AA}$ is sensitive to both spectra shape and path
length dependence, and $I_{\rm AA}$ vs.\ $v_2^{I_{\rm AA}}$ is
sensitive to all three ingredients. 

We found that only our energy loss model with $\Delta E\sim l^3$ AdS/CFT-like energy loss and a CGC medium
describes the $R_{\rm AA}$ vs.\ $v_2$ data well as a function of
centrality and is also qualitatively (within two to three standard deviations)
consistent with the $R_{\rm AA}$ vs.\ $I_{\rm AA}$ data as a
function of the trigger $p_T$. While the $l^2$, pQCD-like energy loss model is
completely inconsistent with the $R_{\rm AA}$ vs.\ $v_2$
correlation, the $l^2$ model describes the $R_{\rm AA}$ vs.\
$I_{\rm AA}$ correlations better than the $l^3$ model, possibly
indicating important physics such as hadronization or flow-coupling
effects missing from our calculation.  (Note that the former is
unlikely given the very large $v_2$ values seen at very large
$p_T\sim10$ GeV/c.)  More dynamical calculations of these
correlations should provide more quantitative and detailed
information, and it is to this end that the proposed correlations
should be useful.

We also identified a number of interesting experimental measurements that---once analyzed---will help further constrain energy loss calculations.  Our absorption model showed a nontrivial change in the centrality dependence of the $R_{AA}$ vs.\ $I_{AA}$ correlation as a function of the momentum of the trigger particle, due to the hardening of the away-side spectrum; currently there is only a single data point for this correlation for any particular trigger $p_T$.  Interestingly, the absorption model predicts a universal eccentricity-scaled correlation between $v_2$ and $v_2^{I_{AA}}$ for both $l^2$ and $l^3$ path length dependencies irrespective of the medium geometry; it would be very interesting to see if this universal relationship is observed in data.  Most fascinating is the new correlation measurement between $I_{AA}$ and $v_2^{I_{AA}}$.  For the given di-hadron suppression, the very large magnitude of the anisotropy of this suppression cannot be described by either $l^2$- or $l^3$-type energy loss models.  Future measurements that reduce the experimental uncertainty on $v_2^{I_{AA}}$ may be extremely difficult to reconcile with current notions of high-$p_T$ energy loss physics.


\section*{Acknowledgements}
This research is supported by the NSF under award number PHY-1019387.
\def\eprinttmppp@#1arXiv:@{#1}
\providecommand{\arxivlink[1]}{\href{http://arxiv.org/abs/#1}{arXiv:#1}}
\def\eprinttmp@#1arXiv:#2 [#3]#4@{\ifthenelse{\equal{#3}{x}}{\ifthenelse{
\equal{#1}{}}{\arxivlink{\eprinttmppp@#2@}}{\arxivlink{#1}}}{\arxivlink{#2}
  [#3]}}
\providecommand{\eprintlink}[1]{\eprinttmp@#1arXiv: [x]@}
\renewcommand{\eprint}[1]{\eprintlink{#1}}
\providecommand{\adsurl}[1]{\href{#1}{ADS}}
\renewcommand{\bibinfo}[2]{\ifthenelse{\equal{#1}{isbn}}{\href{http://cosmologist.info/ISBN/#2}{#2}}{#2}}

\end{document}